\documentclass[12pt,preprint]{aastex}
\usepackage{amsmath}
\usepackage{color}\newcommand{\del}[2]%
{\frac{\mathrm{d}{#2}}{\mathrm{d}{#1}}}
\newcommand{\Del}[2]%
{\frac{\mathrm{D}{#2}}{\mathrm{D}{#1}}}
\newcommand{\ddel}[2]%
{\frac{\mathrm{d}^2{#2}}{\mathrm{d}{#1}^2}}
\newcommand{\pdel}[2]%
{\frac{\partial{#2}}{\partial{#1}}}
\newcommand{\pddel}[2]%
{\frac{\partial^2{#2}}{\partial{#1}^2}}

\newcommand{\cosec}{\mathop{\mathrm{cosec}}\nolimits}

\newcommand{\Ms}{M_{\odot}}
\newcommand{\km}{\,\, \mathrm{km}}

\newcommand{\gauss}{\,\, \mathrm{G}}

\newcommand{\gpcmc}{\,\, \mathrm{g \,\, cm^{-3}}}

\shorttitle{General Relativistic Neutrino Transport in Collapsars}
\shortauthors{Harikae et al.}

\begin{document}

\title{General Relativistic Ray-Tracing Method for Estimating the Energy and Momentum 
 Deposition by Neutrino Pair Annihilation in Collapsars}

\author{Seiji Harikae\altaffilmark{1,2,4}, 
Kei Kotake\altaffilmark{2,3}, Tomoya Takiwaki\altaffilmark{3},
 and Yu-ichiro Sekiguchi\altaffilmark{2}}

\altaffiltext{1}{Department of Astronomy, The Graduate School of Science,
University of Tokyo, 7-3-1 Hongo, Bunkyo-ku, Tokyo, 113-0033, Japan}
\altaffiltext{2}{Division of Theoretical Astronomy, National
  Astronomical Observatory of Japan, Mitaka, Tokyo 181-8588, Japan}
\altaffiltext{3}{Center for Computational Astrophysics, National
  Astronomical Observatory of Japan, Mitaka, Tokyo 181-8588, Japan}
\altaffiltext{4}{Present affilitation: Quants Research Department, Financial Engineering Division, Mitsubishi UFJ Morgan Stanley Securities Co., Ltd., Marunouchi Bldg., 2-4-1, Marunouchi, Chiyoda-ku, Tokyo, 100-6317, Japan}

\email{kkotake@th.nao.ac.jp}

\begin{abstract}
Bearing in mind the application to the collapsar models
 of gamma-ray bursts (GRBs), we develop a numerical scheme and code for 
estimating the deposition of energy and 
momentum due to the neutrino pair annihilation 
($\nu + {\bar \nu} \rightarrow e^{-} + e^{+}$) in the vicinity of accretion tori
 around a Kerr black hole. 
Our code is designed to solve the general relativistic neutrino transfer 
by a ray-tracing method.  To solve the collisional Boltzmann 
equation in curved spacetime, we numerically 
integrate the so-called rendering equation along the null geodesics.
 We employ the Fehlberg(4,5) adaptive integrator in the Runge-Kutta method
 to perform the numerical integration accurately. 
For the neutrino opacity, the charged-current $\beta$-processes are taken into 
account, which are dominant in the vicinity of the accretion tori.
The numerical accuracy of the developed code is certificated by several
 tests, in which we show comparisons with the corresponding analytic solutions. 
 In order to solve the energy dependent ray-tracing transport, we propose that 
an adaptive-mesh-refinement approach, which we take 
for the two radiation angles $(\theta,\phi)$ and the neutrino energy, is useful to 
  reduce the computational cost significantly. Based on the hydrodynamical data in our 
collapsar simulation, we estimate the annihilation rates 
in a post-processing manner.
Increasing the Kerr parameter from 0 to 1, 
 it is found that the general relativistic effect can increase the 
local energy deposition rate by about one order of magnitude, and 
the net energy deposition rate by several tens of percents.
After the accretion disk settles into a stationary state (typically later than 
 $\sim 9$ s from the onset of gravitational collapse),
 we point out that the neutrino-heating timescale in the vicinity of the 
polar funnel region can be shorter than the dynamical timescale. 
Our results suggest the neutrino pair annihilation 
 has a potential importance equal to the conventional magnetohydrodynamic mechanism
  for igniting the GRB fireballs.
\end{abstract}


\section{Introduction}

\label{sec:intro}
Gamma-ray bursts (GRBs) have long attracted the attention of astrophysicists 
 since their accidental discovery in 1970s.
 Regarding the long-duration GRBs, there have been 
 accumulating observations identifying 
 a massive stellar collapse as their origin 
 (e.g., \citet{woos} for a review).
The duration of the long bursts 
may correspond to the accretion of debris falling into the 
central black hole (BH) \citep{piro98}. It suggests the observational 
consequence of the BH formation likewise the supernova of neutron star 
formation. 
For their central engines, 
the so-called collapsar has received quite some interest for more than decade 
\citep{woos93,pacz98,macf99}.

In the collapsar scenario, the central cores 
with significant angular momentum collapse into a black hole. 
Neutrinos emitted from the accretion disk heat matter in the polar funnel region to
launch the GRB outflows.
 \citet{paz90,mezree} pioneerlingly 
proposed that the energy deposition proceeds predominantly
 via neutrino and antineutrino annihilation into electron and positron 
(e.g., $\nu + {\bar \nu} \rightarrow e^{-}+ e^{+}$, hereafter 
``neutrino pair annihilation'').
In addition, it is suggested that the strong magnetic fields in the
cores of order of $10^{15} \gauss$ play also an active role both for driving 
the magneto-driven jets and for extracting a significant amount of
energy from the central engine (e.g.,
\cite{bz77,thom04,uzde07a} and see references therein). 

However, it is still controversial whether the generation of the relativistic outflows 
proceeds predominantly via magnetohydrodynamic (MHD) or neutrino-heating processes. 
So far, much attention has been paid to the MHD processes 
(e.g., \citet{prog03b,mizu04b,lyuti06,fuji06,naga07,mck,komi,barkov,naga09,hari09a}).
A general outcome of these extensive MHD simulations is that the magneto-driven 
shock waves can blow up massive stars along the rotational axis.
  Those primary jet-like explosions are firstly at most mildly relativistic due to 
too much baryons in the central core (e.g., \citet{taki09}), however could 
be relativistic as they propagate further out \citep{naga09}. In such a 
 collapsar environment, explosive nucleosynthesis (e.g., \citet{fuji06,naga07}, 
neutrino and gravitational-wave signals (e.g., \citet{kawa09,hira_05}),
 have been also extensively studied.

In contrast to such blossoms in the MHD studies, 
 there have been only a few studies pursuing the possibility of 
generating jets by the energy deposition via neutrino pair annihilation.
This is mainly because the neutrino emission from 
the accretion disk generally becomes highly aspherical, thus demanding us to 
solve a multidimensional neutrino transfer problem (e.g., \citet{tubbs78,janka89}).
 This is still computationally very expensive, 
 which is also the case for the neutrino-driven supernova simulations 
(see references in \citet{janka07}). For the first time in the collapsar simulations, 
\citet{macf99} pointed out the importance of the energy 
deposition via neutrino pair annihilation, however the energy deposition 
 rates to the polar funnel region were adjusted by hand to produce jets.
To our best knowledge, the fast and collimated neutrino-heated outflows 
have not been realized so far in the numerical simulations  
without the artificial energy injection to the polar funnel regions (see, e.g., 
\citet{aloy00,zhang03,mizuta} and references therein).

Thus far, there have been reported several methods aiming to implement 
the neutrino pair annihilation into the collapsar simulations. 
By estimating the fluxes and spectra of the neutrino emission
 from the accretion disk via the so-called neutrino leakage scheme,
 \citet{ruff97,ruff98} proposed to estimate the heating rate by summing up 
 the contributions of the neutrino and antineutrino radiation incident from 
all directions. 
Along this prescription, \citet{naga07} have estimated the 
 neutrino heating rates, and included them to the hydrodynamical simulation.
  For reducing the computational time, 
 they added one more assumption of the optically thinness 
of the accretion disk to the prescription by \citet{ruff98}.
Even with this potential 
overestimation of the heating rates, the neutrino-driven outflows
 were not observed in their simulations.
More recently, \citet{dess09} have developed a new scheme 
to estimate the energy deposition rate 
using the state-of-the-art, multi-angle neutrino-transport solver \citep{ott08}. 
They discussed the possible formation of the neutrino-driven outflow 
in the postmerger phase of binary neutron-star coalescence.
  Relying on the neutrino leakage scheme,
\citet{hari10} have proposed a special relativistic ray-tracing method to
  estimate the annihilation rates.
Using hydrodynamical data in their
collapsar simulation, they pointed out that the neutrino-heated 
 outflow might be formed in $\sim$ 10 seconds after the initial collapse 
 of the progenitor star.

It should be noted that all of the above schemes neglect 
 the general relativistic (GR) effects for simplicity, 
which have been reported to enhance the annihilation rates 
 significantly near the accreting black holes
 (e.g., \cite{jaros93,jaros96,salm99,asano1,asano2,birkl}).
 Among the GR studies, the numerical method of \citet{birkl} would be one of the most 
sophisticated one, in which a ray-tracing calculation is performed to follow 
the neutrino trajectories in a Kerr spacetime. 
The ray-tracing method has an advantage 
because it can straightforwardly capture important GR features such as the ray bending 
 and redshift. 
 In their scheme, the neutrino number flux emitted
 from the accretion disk (or from the neutrino spheres) is simply assumed to be 
conserved along the geodesics.
In reality, the neutrino emission, absorption, and scattering
 should occur along the neutrino geodesics changing its neutrino distribution 
function simultaneously.
Especially in the absence of the charged-current neutrino interactions, 
the annihilation rates in \citet{birkl} could be overestimated. 
To improve these issues, one has to solve the general relativistic neutrino 
transport equation along each ray, which we are to investigate in this paper.

In this study, we present a numerical code and scheme
for calculating the deposition of energy and momentum via neutrino pair annihilation
 in a Kerr spacetime, in which we 
solve the general relativistic radiative equation along the null geodesics. 
 The charged-current $\beta$-processes are taken into account, which are dominant 
   in the vicinity of the accretion tori (e.g., \citet{dess09}). 
With these improvements, 
the newly developed code would provide a more realistic estimation of the annihilation 
rates than before.
  We check the numerical accuracy of the developed code by showing
 several comparison with analytic solutions, some of 
which we newly derive in this paper. 
 Based on the results of our long-term collapsar simulation \citep{hari09a},
 we run our new code to estimate the annihilation rate in a post-processing manner and
 discuss their implications on the dynamics of collapsars.

This paper is organized as follows. 
In Section \ref{sec:ann},  we summarize the formulation of the 
 general relativistic ray-tracing method for the collisional Boltzmann equation.
Section \ref{sec:test} is devoted to the numerical tests. 
In Section \ref{sec:appl}, we estimate the annihilation rates in a 
post-processing manner using hydrodynamical data in our 
collapsar simulation.
 We summarize our results and discuss their implications in  Section \ref{sec:discussion}. 
 
\section{Neutrino Pair Annihilation in General Relativity}
\label{sec:ann}

\begin{figure}[tbd]
\begin{center}
\includegraphics[scale=0.5]{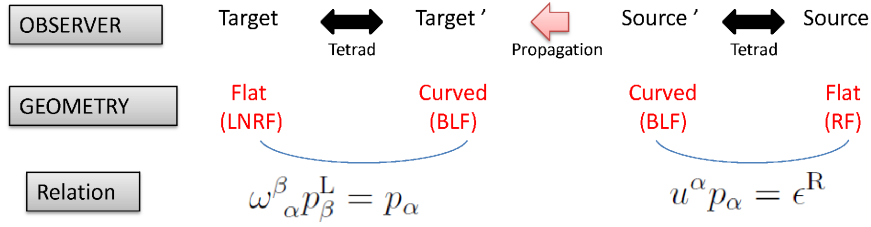}
\caption{Schematic picture of relations of energy/momentum among three frames 
(LNRF, BLF, RF). Upper script R denotes the variables measured in RF, while upper script 
L denotes the variables measured in LNRF.}
\label{fig:observer}
\end{center}
\end{figure}

In this section, we summarize the formalism and our strategy to estimate the 
neutrino-pair-annihilation rates based on the general relativistic radiation transfer.
In section 2.1, we summarize the method to solve the neutrino geodesics 
in a Kerr spacetime for the collisionless Boltzmann equation.
 Then in section 2.2, we move on to mention how to solve the collisional 
Boltzmann equation along the geodesics.

We assume that the gravitational field, which leads to ray bending and redshift,
 is given by the central Kerr BH of mass $M$ and angular momentum parameter
 $a \equiv J/M$ (where $J$ is the angular momentum of the 
BH, and $0 \leq a/M \leq 1$),
 whose metric is given in the Boyer-Lindquist coordinates ($t,r,\theta,\phi$) by
\begin{eqnarray}
ds^2
&=&g_{\alpha\beta}dx^\alpha dx^\beta,\nonumber\\
&=&-\alpha^2 dt^2 +\gamma_{ij}(dx^i+\beta^i dt)(dx^j+\beta^j dt), 
\end{eqnarray}
where the lapse function $\alpha$, the shift vector $\beta^i$ and 
the non-vanishing components of the spatial metric $\gamma_{ij}$ are given as 
\begin{eqnarray}
\alpha=\sqrt{\frac{\Sigma \Delta}{A}},~
\beta^\phi=-\omega,~
\gamma_{rr}=\frac{\Sigma}{\Delta},~
\gamma_{\theta\theta}=\Sigma,~
\gamma_{\phi\phi}=\frac{A\sin^2\theta}{\Sigma},
\end{eqnarray}
where $\Sigma=r^2+a^2\cos^2\theta$, $\Delta=r^2-2Mr+a^2$, 
$A=(r^2+a^2)^2-a^2\Delta\sin^2\theta=\Sigma\Delta+2Mr(r^2+a^2)$, and
$\omega = 2aMr/A$ (e.g., \citet{misner}). 
Here we use $G=c=1$ unit 
and note that the Latin indices ($i,j$) have the domain of
$(r,\theta,\phi)$. For later convenience, we also define 
the dimensionless angular momentum parameter of $a^* = a/M$.

For later convenience, we first introduce the following three frames,  
 the Boyer-Lindquist frame (BLF) which is given by the center of mass 
system in curved space-time, the locally non-rotating frame (LNRF) 
 which is given by the tetrad frame rotating with the central BH to make 
  the dragging effects vanish (i.e., $e_i = 0$ 
 with $e_{\mu}$ being the basis of the vierbein),
and the rest frame of fluid (RF) which is necessary to define 
quantities related to radiation such as emissivity and absorptivity. 
These three frames can be connected with each other
 by the tetrad and Lorentz transformations. In the following sections, 
 the quantities measured in the LNRF and RF are denoted by the superscript ``L''
 and ``R'', respectively. Variables in the BLF are denoted without 
any superscripts. A schematic picture between these three frames are illustrated 
in Figure \ref{fig:observer}. What we finally need is the annihilation rates
 measured by the observer in the LNRF (left-end in the figure). The neutrino
 emissivity and absorptivity are naturally defined in the rest frame of fluid (RF), 
in which the radiation isotropy is maintained (right-end in the figure). The first 
step is to transform the variables in the RF to the ones in the BLF 
(from the right-end to left by one step) 
using the tetrad transformation (the ``relation'' shown in the figure with some 
 formulae will be derived later in this section). 
The second step is to do the ray-tracing calculation 
 from the source to the target in the global BLF (indicated by ``Target$'$ '' in 
 the figure). Finally the annihilation rates in the LNRF is given by the tetrad 
 transformation form the BLF to the LNRF. In the following, we explain these 
procedures more in detail.

The local annihilation rate in the LNRF 
(e.g., \citet{good87,asano2,birkl}) is written as, 
\begin{eqnarray}\label{eq:heatdefgr}
    Q_{\mu}^{\rm L} (\mbox{\boldmath$r$})&=&
    2 K G_{\rm F}^2 
    \int
    d^3 {\mbox{\boldmath$p$}}^{\rm L}_\nu d^3 {\mbox{\boldmath$p$}}^{\rm L}_{\bar \nu} \nonumber \\
    && \times (\epsilon^{\rm L}_{\nu} \epsilon^{\rm L}_{\bar{\nu}}) 
    ({\mbox{\boldmath$p$}}^{\rm L}_\nu + {\mbox{\boldmath$p$}}^{\rm L}_{\bar{\nu}} )_{\mu} 
	f^{\rm L}_{\nu}(\mbox{\boldmath$p$}^{\rm L}_{\nu}, \mbox{\boldmath$r$})
    f^{\rm L}_{\bar{\nu}}(\mbox{\boldmath$p$}^{\rm L}_{\bar{\nu}}, \mbox{\boldmath$r$})
    \nonumber \\
    && \times \left[ 1-\sin{\theta_{\nu}} \sin{\theta_{\bar{\nu}}}
        \cos{(\varphi_{\nu}-\varphi_{\bar{\nu}})}
        -\cos{\theta_{\nu}} \cos{\theta_{\bar{\nu}}}
    \right]^2, 
\end{eqnarray}
where $f^{\rm L}_\nu$ is the number density of neutrinos in the phase space 
 within the solid angle of 
$d\Omega_{\nu} = \sin{\theta_{\nu}}d\theta_{\nu}d\varphi_{\nu}$ 
in the momentum space, 
$\mbox{\boldmath$p$}^{\rm L}_\nu$ and $\epsilon^{\rm L}_\nu$ is the momentum and 
energy in the LNRF, respectively. 
Those definitions are the same for antineutrino by changing the notation 
$\nu$ to ${\bar \nu}$.
The dimensionless parameter $K$ is written as 
\begin{eqnarray}
K(\nu_e, {\bar \nu}_e) = \frac{1+4\sin^2\theta_{\rm W}+8\sin^4\theta_{\rm W}}{6 \pi}, \\ 
K(\nu_\mu, {\bar \nu}_\mu) = K(\nu_\tau, {\bar \nu}_\tau) = \frac{1-4\sin^2\theta_{\rm W}+8\sin^4\theta_{\rm W}}{6 \pi}.  
\end{eqnarray}
Here the Fermi constant is $G_{\rm F}^2=5.29 \times 10^{-44} {\rm cm^2\, MeV^{-2}}$ and
the Weinberg angle is $\sin^2{\theta_{\rm W}}=0.23$. 
Since the ray-tracing calculation is conveniently done 
 in the global BLF, we transform the neutrino momentum in the LNRF ($p^{\rm L}_{\alpha}$)
 to $p_{\alpha}$ measured in the BLF. 
This transformation is done by the tetrad transformation as,
\begin{equation}
	p_{\alpha} = {\omega}^{\beta}_{~\alpha} p^{\rm L}_{\beta},
\end{equation}
where ${\omega}^{\beta}_{~\alpha}$ is the transformation matrix of the 
Boyer-Lindquist coordinates,
\begin{equation}
\omega^\beta_{~\alpha} =
  \begin{pmatrix}
   \alpha & 0 & 0 & 0 \\
   0 & \sqrt{\gamma_{rr}} & 0 & 0 \\
   0 & 0 & \sqrt{\gamma_{\theta \theta}} & 0\\ 
   \beta^\phi \sqrt{\gamma_{\phi \phi}} & 0 & 0 & \sqrt{\gamma_{\phi \phi}}\\
  \end{pmatrix}. 
\end{equation}
 It is noted that the annihilation rate in the BLF is 
 given as,
\begin{equation}
	Q_{\mu} = {\omega}^{\nu}_{~\mu} Q_{\nu}^{\rm L},
\end{equation}
 which can be readily implemented in the general relativistic hydrodynamic simulations
 via $\nabla_{\nu} T^{\mu\, \nu} = Q^{\mu}$
(e.g., \citet{shiba}), although this is beyond the scope of this paper. 

To evaluate the annihilation rates, we have yet to determine 
$f^{\rm L}_{\nu ({\bar \nu})}$ 
in Equation (\ref{eq:heatdefgr}). It is noted that 
the distribution function is invariant under the tetrad transformation as 
\begin{equation}
f^{\rm L}_{\nu ({\bar \nu})}(\mbox{\boldmath$p$}^{\rm L}_{\nu ({\bar \nu})}, \mbox{\boldmath$r$}) = f_{\nu ({\bar \nu})}(\mbox{\boldmath$p$}_{\nu ({\bar \nu})}, \mbox{\boldmath$r$}), 
\end{equation}
where $f_{\nu ({\bar \nu})}$ 
is the distribution function in the BLF. 
 $f_{\nu ({\bar \nu})}$ is determined by the general relativistic Boltzmann transport 
equation \citep{mis64} as 
\begin{eqnarray}
\frac{df_{\nu ({\bar \nu})}}{d\lambda} &=&  p^{\alpha}\frac{Df_{\nu ({\bar \nu})}}{Dx^{\alpha}} = \left( \frac{df_{\nu ({\bar \nu})}}{d\lambda} \right)_{\rm coll}, \label{eq:rt} \\
\frac{D}{Dx^{\alpha}} &\equiv& \frac{\partial}{\partial x^{\alpha}} 
- \Gamma_{\alpha \gamma}^{\beta}p^{\gamma}\frac{\partial}{\partial p^{\beta}}, \label{eq:geo}
\end{eqnarray}
where $\left( df_{\nu ({\bar \nu})} / d\lambda \right)_{\rm coll}$ represents 
the collision term.
 In the context of photon propagation from the accretion disk,
 there have been extensive studies to determine the geodesics
(e.g., \citet{car68,bar72,cun73,cun75,rauch94,fanton97,cad98,cad03,cad05a,
cad05b,lili05,mull04,taka04,taka05,taka07}). 
Since the mass of neutrinos are negligible compared to the 
relevant energy-scales to affect the dynamics of collapsars ($O({\rm MeV})$), 
the neutrino geodesics can be treated as that of photon and the techniques for the 
 photon transfer is also applicable to neutrinos. 
To determine the null geodesics,
 we basically follow the method in \citet{zink08} which utilizes the ray-tracing method.
To treat the neutrino transport equation with the collision term, 
we employ the formalism 
 developed by \citet{lindq66}. In the following, we summarize the method to 
 determine the geodesics in a Kerr spacetime for the collisionless Boltzmann equation.
 Then in section 2.2, we present the method to solve the collisional Boltzmann equation.

\subsection{Geodesics in a Kerr Geometry}
\label{sec:geo_kerr}

In the Boyer-Lindquist coordinates, the Lagrangian ${\cal L}$ for describing the geodesics of massless particles in the Kerr 
 geometry  (e.g., \citet{misner})
 is given as 
\begin{eqnarray}
	2 {\cal L} &\equiv& g_{\alpha \beta} \frac{dx^{\alpha}}{d\lambda} \frac{dx^{\beta}}{d\lambda} \nonumber \\
	&=& - \left(1-\frac{2Mr}{\Sigma}\right)\dot{t}^2 
      - \frac{4aMr\sin^2 \theta}{\Sigma}\dot{t}\dot{\phi} 
      + \frac{\Sigma}{\Delta}\dot{r}^2 \nonumber \\
      &&+ \Sigma \dot{\theta}^2
      + \left[ r^2+a^2+\frac{2a^2 Mr\sin^2 \theta}{\Sigma} \right] \sin^2 \theta~\dot{\phi}^2, \label{eq:lag}
\end{eqnarray}
where overdots denote the differentiation with respect to an affine parameter 
$\lambda$.  With three constants of motion,
\begin{eqnarray}
	 E &\equiv& -p_{t}, \\
	 L_{z} &\equiv& p_{\phi}, \\
	{\cal C} &\equiv& \left[L_{z}^2 \cosec^2 \theta -a^2E^2 \right] \cos^2\theta + p_{\theta}^2,
\end{eqnarray}
one obtains equations governing the orbital trajectory (e.g., \citet{car68,bar72}),
\begin{eqnarray}\label{eq:conmom_stat}
	p^{t     } &=& \frac{E - \omega L_{z}}{\alpha^2}, \\
	p^{r     } &=& {\rm sign} \left(\frac{dr     }{d\tau}\right) \frac{\sqrt{\cal R}}{\Sigma}  , \\
	p^{\theta} &=& {\rm sign} \left(\frac{d\theta}{d\tau}\right) \frac{\sqrt{\Theta}}{\Sigma}  , \\
	p^{\phi  } &=& \frac{\frac{L_{z}}{\sin^2\theta}\left( 1-\frac{2Mr}{\Sigma} \right) + \frac{2Mr}{\Sigma}aE }{\Delta},  
	\label{eq:conmom_end}
\end{eqnarray}
where 
\begin{eqnarray}
	{\cal R} &\equiv& {\cal P}^2 - \Delta [(L_z - aE)^2 + {\cal C}], \\
	 \Theta  &\equiv& {\cal C} - [-a^2 E^2 + L_z^2 \cosec^2 \theta ] \cos^2 \theta, \\
	{\cal P} &\equiv& (r^2+a^2)E - aL_{z}. 
\end{eqnarray}
 The integrals of motion in Equations (\ref{eq:conmom_stat}) 
- (\ref{eq:conmom_end}) can be performed either numerically or analytically.
Although the analytic solutions, if obtained, are accurate and good for reducing the
 computational cost of the ray-tracing calculation, they may be obtained only 
for some special conditions such as the motion for $\phi = 0$ 
(in the $(r,\theta)$ plane).
Therefore we choose to perform the 
 numerical integration, and utilize the analytic solutions to test the 
 validity of the numerical integration in some test problems that will be presented in 
 section 3.


To capture accurately the trajectory in the vicinity of the BH, a much finer resolution 
 with respect to $\lambda$ should be taken than for the regions far distant 
 from the BH. Therefore some adaptive-mesh-refinement approach is needed for 
 accurate and efficient numerical integration.
 As in \citet{zink08}, we choose to employ the scaled fourth-order Runge-Kutta method
  (\citet{Fehl70}, see also \citet{papa88}), which is often referred 
 to as the RKF45 method. 
  Our choice of the Fehlberg (4,5) adaptive integrator
 is known to be very useful because it is possible to estimate 
a truncation error, by which the adequate step-sizing for the Runge-Kutta
 integration can be determined automatically.
In this method, the step size in each integration 
is controlled by comparing the residual error 
$d_{k}^{\,\alpha}$ to a given criterion $\delta$ at every $k$ step 
(see \citet{Fehl70} for more detail). 
Here we set $\delta = 10^{-4}$, which provides enough accuracy in 
tracing the ray near the BH, as will be shown in section 3. 

\subsection{Radiative Transfer in Curved Space-time}
\label{sec:rt}

According to \citet{lindq66}, the Boltzmann equation with the collision term for photons
 is 
 generally expressed as,
\begin{equation}
 \frac{df}{d\lambda} = n (Q - \kappa f).
\label{render}
\end{equation}
Here $n(\mbox{\boldmath $x$})$ is the proper number density of the external 
 medium with which neutrinos interact, and thus measured in its own local rest 
frame. $Q(\mbox{\boldmath $x$},\mbox{\boldmath $p$})$ is the emission rate per particle
 of the medium ($Q_{\rm e}$), plus a further increase due to scattering 
 ($Q_{\rm s}$), which can be therefore  
 written as 
\begin{eqnarray}
Q(\mbox{\boldmath $x$},\mbox{\boldmath $p$})         &=& Q_{\rm e}(\mbox{\boldmath $x$},\epsilon^{\rm R})
 + Q_{\rm s}(\mbox{\boldmath $x$},\mbox{\boldmath $p$}), \\
Q_{\rm e}(\mbox{\boldmath $x$},\epsilon^{\rm R}) &=& \frac{j(\mbox{\boldmath $x$},\epsilon^{\rm R})}{4\pi (\epsilon^{\rm R})^2}, \label{absorption}\\
Q_{\rm s}(\mbox{\boldmath $x$},\mbox{\boldmath $p$}) &=& 
\int {\epsilon '}^{\rm R} d{\epsilon '}^{\rm R}
d\Omega (\mbox{\boldmath $x$},\mbox{\boldmath $p$}') 
\xi (\mbox{\boldmath $x$}; \mbox{\boldmath $p$}' \rightarrow \mbox{\boldmath $p$}) 
f(\mbox{\boldmath $x$},\mbox{\boldmath $p$}'), \label{scatt}
\end{eqnarray}
 where $j$ is the emissivity and $\xi (\mbox{\boldmath $x$}; \mbox{\boldmath $p$}' 
\rightarrow \mbox{\boldmath $p$})$ is the so-called invariant phase function, describing 
 the momentum transfer due to scattering. $\kappa$ in Equation (\ref{render}) 
is the invariant absorption coefficient. 
$d\Omega (\mbox{\boldmath $x$},\mbox{\boldmath $p$})$ is the solid angle 
in the momentum space of $\mbox{\boldmath $p$}$ at position $\mbox{\boldmath $x$}$. 
 $\epsilon^{\rm R}$ is the neutrino energy measured 
in the local proper frame that is related to the quantities in the BLF as 
\begin{eqnarray}
	\epsilon^{\rm R} &=& -p_{0}^{\rm R} \nonumber \\
	&=& -u^\alpha p_{\alpha}. 
\end{eqnarray}

The formal solution of Equation (\ref{render}) can be given as,
\begin{equation}\label{eq:rt3}
	f(\epsilon, \Omega) = \int^{\lambda_{S}}_{\lambda_{0}} n(\lambda'') 
	Q(\lambda'', f)  e^{-\int^{\lambda_{S}}_{\lambda''} n(\lambda') 
	\kappa(\lambda') d\lambda' } d\lambda'', 
\end{equation}
 which is referred to as 
the rendering equation of the radiation transport problem (e.g., \citet{zink08}).
 Note that the integration with respect to $\lambda$ 
 starts from a given target point ($\lambda_{0}$) 
where the neutrino pair annihilation occurs, propagated backward to the neutrino sources
 along the geodesics. This backward ray-tracing terminates when
 it hits the most outer boundary of our computational domain or when the optical depth
 for each neutrino energy exceeds unity indicating the surface of the neutrino spheres, 
both of which are represented by 
$\lambda_S$ in Equation (\ref{eq:rt3}).
 It is noted that we set the inner boundary of the target region to be 
 the surface of the ergosphere, because 
we consider an idealized situation that the energy released inside the ergosphere
will terminate in the BH, playing no important role to energetize a GRB.

In solving the rendering equation, we neglect the scattering terms $Q_{\rm s}$ 
 (Equation (\ref{scatt})), which 
is not only difficult to be treated by the ray-tracing technique but also 
a major undertaking in the radiative transport problem in general. 
The integration in the rendering Equation (\ref{eq:rt3}) 
is done explicitly along the geodesics.
In doing so, we determine each integration step by restricting the maximum change 
of neutrino opacity
 for all the neutrino energy-bins to be less than 10 \%. 
By this choice, our code can safely pass some test problems (see section 3).

Neglecting the energy and momentum transfer via neutrino scattering, 
the neutrino Boltzmann equation for $\nu_e$ and $\bar{\nu}_e$ now reads,
\begin{equation}
 \frac{df}{d\lambda} = n [Q_e(1-f) - \kappa f] = n[Q_e - \kappa^{*} f],
\label{render_neutrino}
\end{equation}
 where the Pauli blocking term:$(1-f)$ is now taken into accout.
 It is noted that the rendering equation is also valid in this case by 
 replacing $\kappa$ in Equation (\ref{eq:rt3}) with 
$\kappa^{*} \equiv (Q_e + \kappa)$.
As for the opacity sources of neutrinos ($\kappa^{*}$), 
electron capture on proton and nuclei, 
positron capture on neutron, neutrino scattering with nucleon and nuclei,
 are included \citep{full85,taka78,bruenn}. Here $\kappa^{*}$ is
 estimated as $\kappa^{*} = \Sigma[n_{\rm target}\cdot \sigma(\epsilon^{\rm R})]$ with 
 $n_{\rm target}$, $\sigma(\epsilon^{\rm R})$ being the target number density of each 
reaction and the corresponding cross section, respectively.
 The neutrino emission illuminated from the accretion disk mainly comes from the 
 optically thick region, where the charged current $\beta$-equilibrium should be 
 nearly satisfied.  Hence we estimate the neutrino emissivity 
as $Q_e = \kappa^{*} f^{\rm FD}$, where $f^{\rm FD} [= 1/(e^{\epsilon^{\rm R}/T}+1)]$
  is the Fermi-Dirac neutrino distribution function with a vanishing chemical potential.

It is noted that an adaptive-mesh-refinement (AMR) approach that 
we propose in this paper, is an another important tool for saving 
the computational cost of the ray-tracing 
 calculation. For example, a number of rays are required for estimating the 
annihilation rates correctly in the vicinity 
 of the accretion disk, in which the neutrino-heated outflow is expected 
to be produced. 
 It is therefore of primary importance to do AMR with respect to the angular 
 direction of rays. Secondly, the energy bin of neutrinos 
 is better to be treated by AMR, because the neutrino 
distribution function can be more accurately determined if the 
finer energy-bins are cast for the relevant energy scales. The actual implementation 
 procedure is given as follows. Given a point $\mbox{\boldmath $x$}$, we search
 the maximum intensity $I(\epsilon, \theta, \phi)$ among the neighboring 
 points for all the direction and for all the energy bins and call it 
as $I_{\rm max}$. Then we focus on the energy bins and angular directions, which
 satisfy $I_{\rm crit} (\epsilon,\theta,\phi) \ge {\cal K} I_{\rm max}$ where 
 we set ${\cal K} = 0.01$.  Only for the domain of $(\epsilon,\theta,\phi)$
satisfying the condition, we cast finer mesh points. In the actual implementation, 
 we perform this selecting procedure for every 3-dimensional space, which 
 merits not only for saving the computational costs but also for maintaining the 
 good accuracy to estimate the annihilation rates. 

\section{Numerical Tests} 
\label{sec:test}

Before applying the newly developed code to collapsars, 
we shall check the accuracy of our code.
In sections \ref{sec:geo_rt} and \ref{sec:geo_rp}, we show a comparison of 
 the neutrino trajectory between the numerical and analytic solution,
 by which we check the numerical accuracy
 to solve the collisionless Boltzmann. In 
 section \ref{sec:bhshadow}, we demonstrate capability of our code 
 to capture the imaging around the accreting black holes, that is the so-called 
  BH shadow problem. In case of the collisional Boltzmann equation, we
 perform the numerical tests to reproduce the radiation fields shedding 
  from a spherical light-bulb, which will be presented in section 3.4.

\subsection{Geodesics in the ($r$-$\theta$) plane}
\label{sec:geo_rt}

\begin{figure}[htb]
\begin{center}
\includegraphics[scale=0.7]{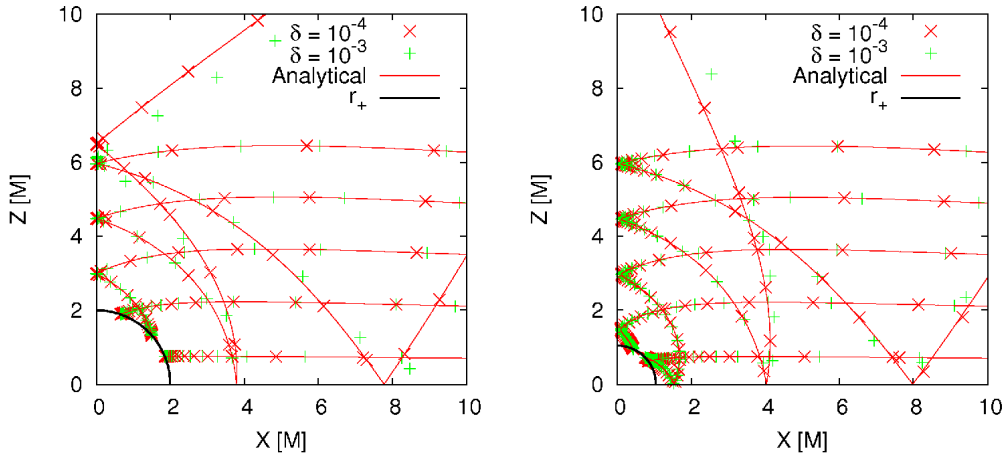}
\caption{Geodesics of neutrinos in the ($r$-$\theta$) plane for the 
dimensionless Kerr parameter 
$a^*=0$ (left) and $a^*=0.999$ (right), obtained either
 analytically (lines) or numerically (points) with the two different 
 regulation parameters ($\delta = 10^{-4}$ (cross) and $10^{-3}$ (plus)) 
for the Runge-Kutta integration (e.g., section 2.1).
 The central black circle (quadrant) represents the event horizon of the
 BH. It is noted that the reflection at $X=0$ or $Z=0$ is just for a 
visualization. At $X=0$ for example, the rays continue to propagate left 
($X < 0$) in reality. }
\label{test_Z}
\end{center}
\end{figure}

By a straightforward, albeit tedious calculation, one can 
 obtain the well-known analytic form of the null 
geodesics in the ($r$-$\theta$) plane around 
 a Kerr BH (e.g., \citet{car68,bar72,cad98,lili05}). 
Figure \ref{test_Z} shows the geodesics near the BH in the case of $a^* = 0$ 
 (left) or $a^*=0.999$ (right), obtained either numerically (points) or analytically 
(lines). In both cases, neutrinos are initially injected from the right edge of the 
figure with different impact parameters (for different $Z$ in the figure). 
They are shown to be 
dragged by the gravity of the BH, whose surface is indicated by the black line 
in the center. 
It is noted that the reflection at $X=0$ or $Z=0$ is just for a 
visualization. For example at $X=0$, the rays keep on propagating to the left ($X < 0$) 
in reality. For the numerical solutions, we vary the two different parameters 
($\delta = 10^{-3}, 10^{-4}$), which regulate the numerical convergence in the 
adaptive integrator (see section \ref{sec:geo_kerr}). 
 We find that the regulation parameter of 
$\delta = 10^{-4}$ is sufficient to trace the trajectory in a good agreement with 
 the analytic solution, which we take in the following calculations.

\subsection{Geodesics in the ($r$-$\phi$) plane}
\label{sec:geo_rp}
\begin{figure}[tb]
\begin{center}
\includegraphics[scale=0.7]{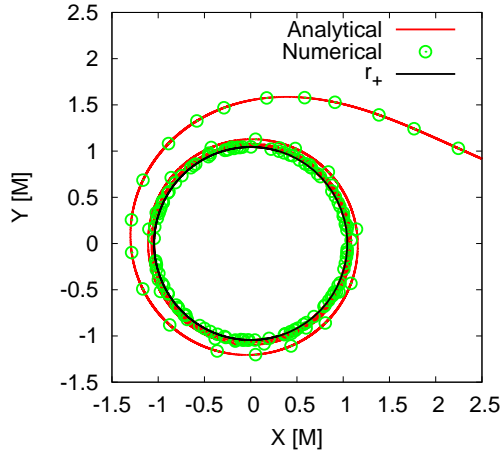}
\caption{Same as Figure \ref{test_Z} but for the numerical solution (circle) 
in the (r-$\phi$) plane. $r_{+}$ indicates the position of the outer event 
horizon. } 
\label{test_rphi}
\end{center}
\end{figure}
Now we move on to show the geodesics in the ($r$-$\phi$) plane. 
Since the analytical solution becomes very complicated in this case, 
we consider a special case that is $L=aE$ \citep{cha83}. 
In this case, the evolution equations (Equations (\ref{eq:conmom_stat}) 
- (\ref{eq:conmom_end})) are greatly simplified as 
\begin{eqnarray}
	{\dot r   } &=& \pm E, \\
	{\dot \phi} &=& \frac{aE}{\Delta}.
\end{eqnarray}
Combining these equations, the geodesics in the ($r$-$\phi$) plane becomes 
\begin{eqnarray}
	\frac{d\phi}{dr} &=& \pm \frac{a}{\Delta}, \\
	\pm \phi &=& \frac{a}{r_{+} - r_{-}} \log \left( \frac{r}{r_{+}} - 1 \right)
			 - \frac{a}{r_{+} - r_{-}} \log \left( \frac{r}{r_{-}} - 1 \right), 
\end{eqnarray}
where $r_{\pm}$ is the position of the event horizon. 

Figure \ref{test_rphi} is the same as Figure \ref{test_Z}, 
but for the geodesics in the ($r$-$\phi$)
 plane around 
an extremely rapidly rotating BH of $a^*=0.999$ (note again 
that $a^* = a/M$ is the dimensionless Kerr parameter). In the following, 
 we call the case of $a^*=0.999$ as an extreme Kerr for simplicity.
 As shown, our 
 numerical integration can reproduce the analytical solution without 
visible errors. These results support that our code can trace correctly
 the null geodesics in the Kerr geometry.

\begin{figure}[htb]
\begin{center}
\includegraphics[scale=0.4]{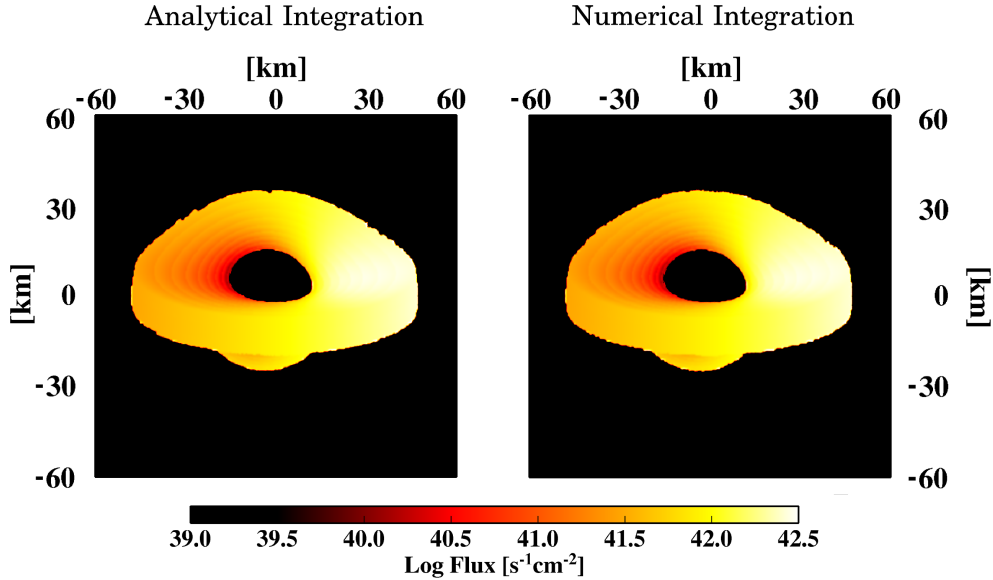}
\caption{Comparison of the neutrino images around the accreting black holes with 
$a^*=0.999$, 
obtained either from the analytic (left) or numerical (right) 
integration of the geodesics. The neutrino flux shown is for the neutrino energy of 
$20 {\rm MeV}$.} 
\label{fig:image_comp}
\end{center}
\end{figure}

\subsection{BH Shadow for Neutrinos}
\label{sec:bhshadow}
In this section, we demonstrate capability of our code
 to capture the imaging around the accreting black holes, which is often 
 referred to as the BH shadow problem.

As for the neutrino sources, we assume a thin accretion disk with a Keplerian rotation profile. We set the mass of the BH to be $2\Ms$ surrounded by the accretion disk, 
whose inner and outer radius are set to be the last stable orbit of the black hole
($r_{\rm lso}$) and $15GM/c^2$ with the disk thickness of $\pi/10$ (rad), respectively.
 The accretion disk is set to have a uniform density, temperature, and electron fraction of $10^{13}\gpcmc$, $5\times10^{11}{\rm K}$, 
and 0.3, respectively. We focus only on the electron-type neutrino in 
 this test problem.

Figure \ref{fig:image_comp} shows one example of the neutrino images around the 
accreting black holes seen from the viewing angle of $\theta_{\rm view} = 72^{\circ}$ 
from the spin axis of the accretion disk.  
No visible differences are seen between the two panels, in which left and right panels 
are obtained either from the analytic or numerical integration of the geodesics.
 This supports the validity of our numerical 
integration of the rendering equation (Equation (\ref{eq:rt3})).
 
\begin{figure}[htb]
\begin{center}
\includegraphics[scale=0.4]{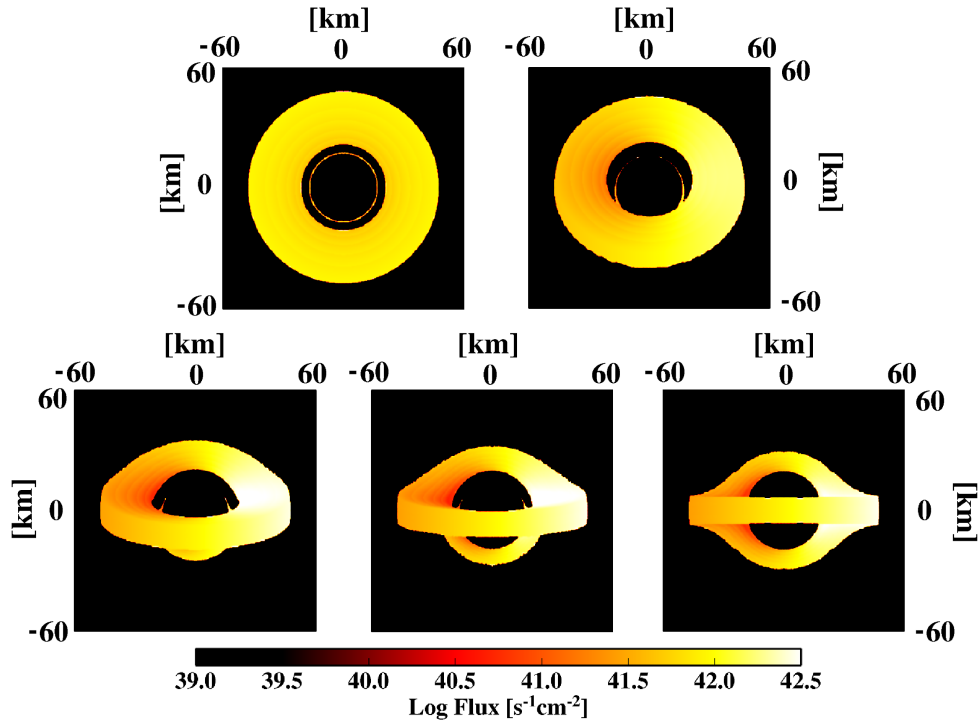}
\caption{Neutrino images of the accretion disk for different viewing angles
 ($\theta_{\rm view}$ = $0^{\circ}$ (top left), $36^{\circ}$ (top right), from 
$72^{\circ}$, $81^{\circ}$ to $90^{\circ}$ (from bottom left to right).
 It is noted that $\theta_{\rm view}$  is the angle measured from the spin axis 
 of the accretion disk. The spin parameter of central BH is set to be $a^* = 0$.} 
\label{fig:image_angle}
\end{center}
\end{figure}

\begin{figure*}[htb]
\begin{center}
\includegraphics[scale=0.4]{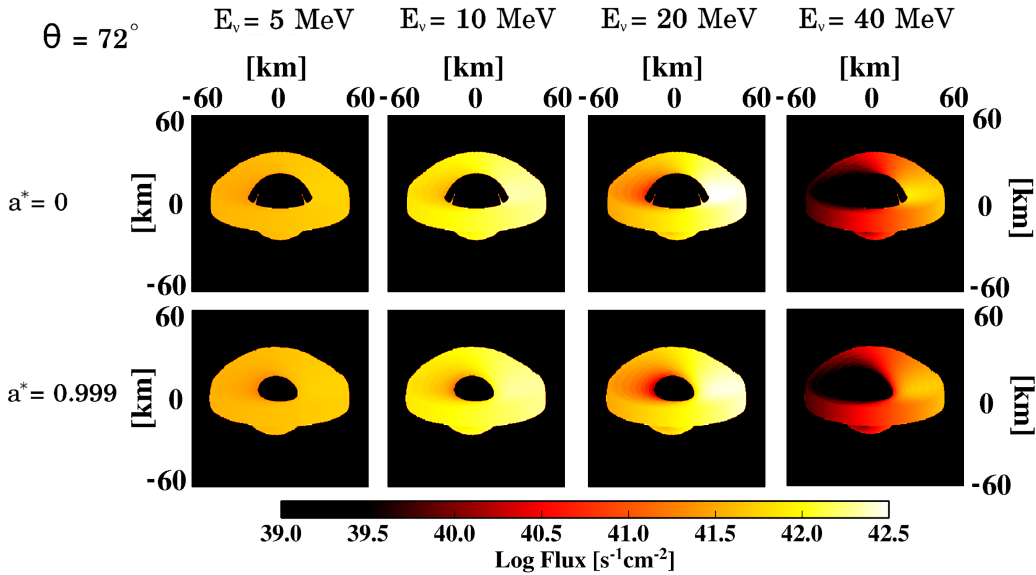}
\caption{Images of the accretion disk for different neutrino energies ($E_{\nu}$) 
and BH spin parameters. The side length of each plot is 60 km.}
\label{fig:bhshadow_a}
\end{center}
\end{figure*}

Figure \ref{fig:image_angle} shows a variety of the images seen from various 
viewing angles. For example, when we see the accretion disk from the equatorial 
plane (bottom right), we can observe neutrinos not only from the disk of the front 
side, but also from the opposite side because of the bending of the trajectory. 

Figure \ref{fig:bhshadow_a} shows the images for different neutrino energies,
 while the viewing angle is kept fixed ($\theta_{\rm view} = 72^{\circ}$).
 For lower energy neutrinos (such as for $5$ MeV (left panel)), 
the disk luminosity is shown to be almost north-south symmetric, 
while it becomes highly asymmetric for higher energy neutrinos (such as 
 for $40$MeV (right panel)). 
As the neutrino energy becomes lower, the position of
 the neutrino sphere is formed deeper inside 
the accretion disk, by which we can  
see the regions closer to the BH
(Figure \ref{fig:bhshadow_a}). Since the angular velocity of the Keplerin disk 
is larger for the distant region from the center, the 
 deformation of the images due to the Doppler effects can be more remarkably seen for 
the high energy neutrinos.
 It is interesting to note that in the case of the maximally rotating black hole
 (bottom panels), the BH shadow becomes asymmetric even for the low energy neutrinos 
due to the frame-dragging effects (bottom two left panels). Such features 
 for the photon shadow in the vicinity of massive BHs in our Galactic center,
 have been considered to
 give an important information to reveal the mass and spin of the BHs
 (e.g., \cite{taka07,nagakura}). 
Although this may not be the case for GRBs due to their cosmological distances, the bending of neutrinos
 may have impacts on the gravitational radiation generated by anisotropic neutrino 
emission (e.g., \citet{epstein,kotake_09}). This can be one 
possible extension of this study.

\begin{figure}[htb]
\begin{center}
\includegraphics[scale=0.6]{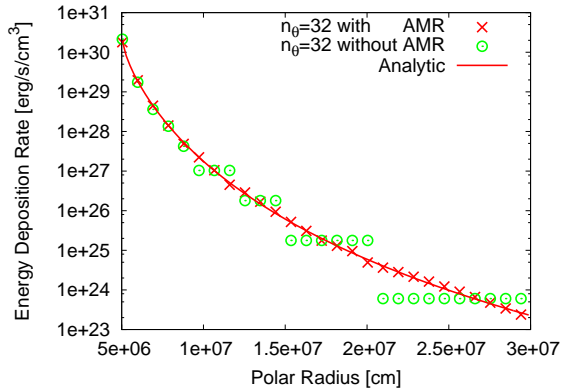}
\caption{Comparison of the energy deposition rate from the spherical light-bulb test 
 (see text for detail) for a given angular resolution of ray-tracing 
calculation ($n_{\theta}$ = 32) with or without the AMR treatment.  
Note in this test that we assume the Minkowskian geometry.}
\label{fig:ann_amr}
\end{center}
\end{figure}

\begin{figure}[hbt]
\begin{center}
\includegraphics[scale=0.6]{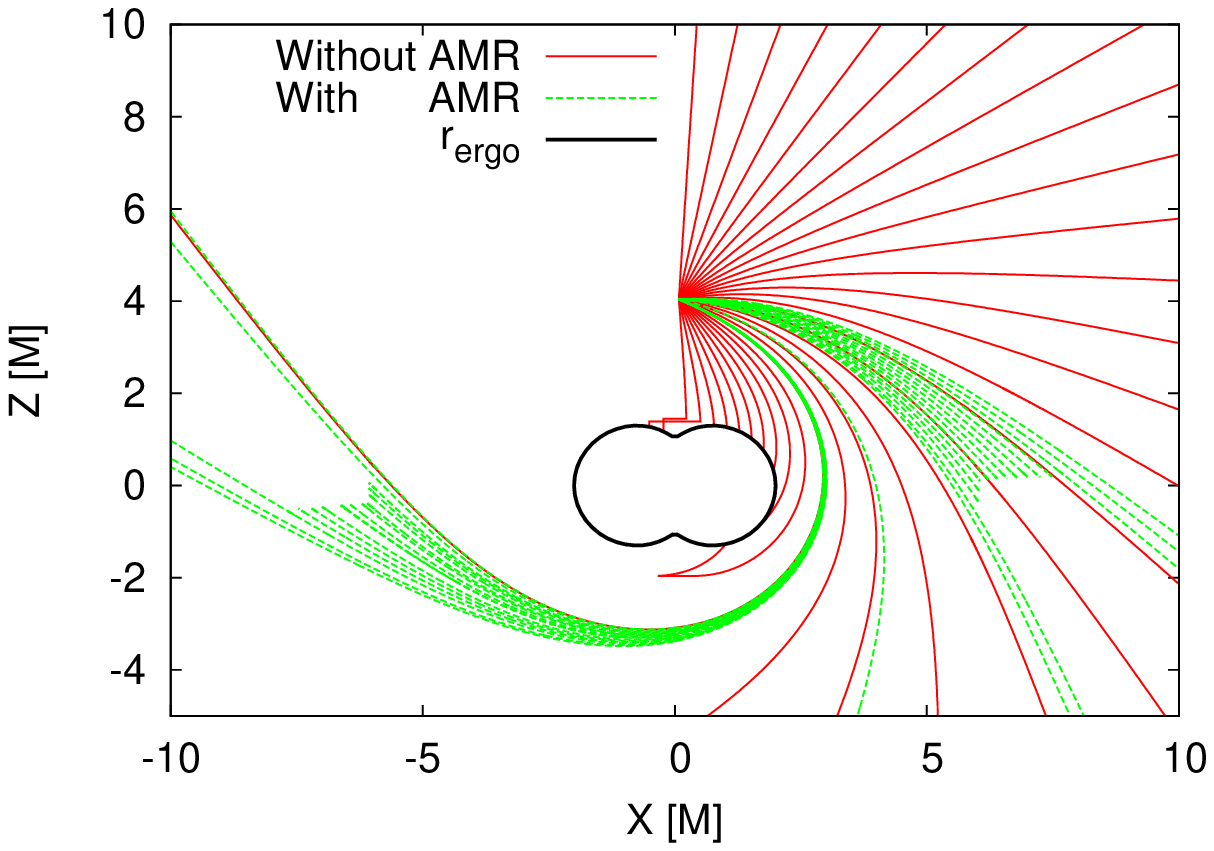}
\caption{One example describing the casting of rays with (green) or without AMR 
technique (red). In this case, the rays are cast for estimating the 
 annihilation rate at a given point (seen as a convergent point of the rays) 
outside the ergosphere ($r_{\rm ergo}$).
 The concentration of the rays is seen (green), which is helpful to correctly
 estimate the heating rates with reduced computational cost. 
Note in this figure that only selected rays are chosen for 
illustrative purpose. This plot is selected from the BH shadow problems 
(section \ref{sec:bhshadow}) for the visualization of the AMR.}
\label{fig:amr2d}
\end{center}
\end{figure}

\subsection{Neutrino Pair Annihilation from a Spherical Neutrino Sphere}
\label{sec:test_ann}

For the collisional Boltzmann equation in GR, it is commonly not trivial to derive
 analytic solutions for a radiative transport problem.
In the following, we derive the analytic solution for radiation fields, 
shedding from a spherical light-bulb into a uniform medium outside.
 We hope that the analytic solution may be useful to check newly developed
 codes for the radiative transport in curved space.

In the following numerical tests, the spherical neutrino sphere with a 
 radius of $50\km$ is assumed to have its surface temperature of $T = 5$ MeV on which the neutrino 
distribution function takes a Fermi-Dirac shape
 with vanishing chemical potential. The numerical domain [$50 \km : 300\km$] is 
covered with $n_r = 100$ radial mesh points. The fiducial values of the energy and 
angular bins for the ray-tracing calculation 
are set to be $(n_{\epsilon^{\rm L}},n_{\theta},n_{\phi}) 
= (16,32,16)$, which we will change to see the numerical convergence.

\begin{figure}[htb]
\begin{center}
\includegraphics[scale=0.3,angle=-90]{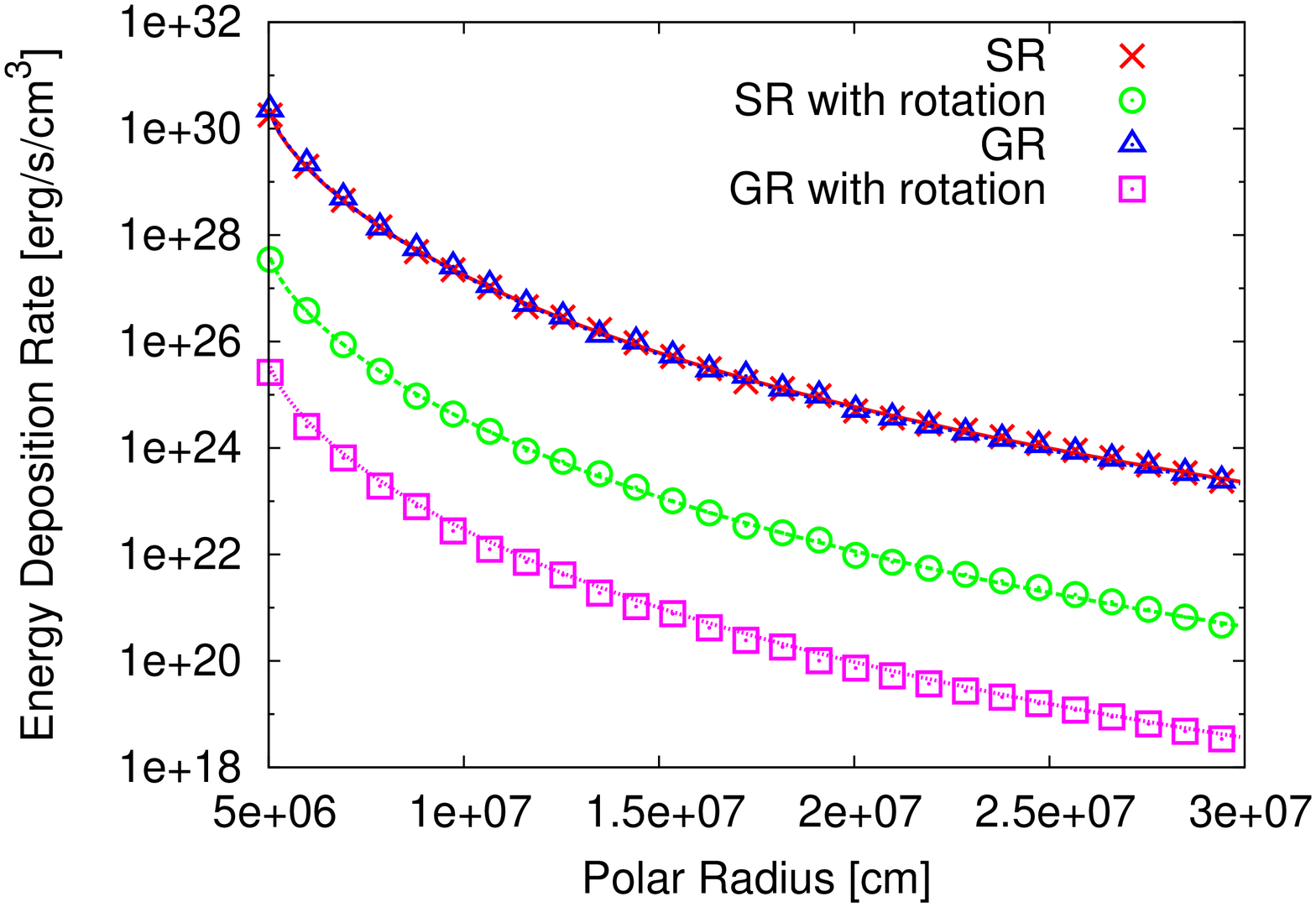}
\caption{Same as Figure \ref{fig:ann_amr} but for 
the Minkowskian case with or without rotation (circle, cross), 
 the Schwarzschild case with or without rotation (square, triangle).
See text for detail. For models with rotation, we set $1/\sqrt{1- ({\hat v}^3)^{2}} = 2$. 
For models with the Schwarzschild geometry, we put a point mass of $M=3\Ms$ inside 
 the neutrino sphere.} 
\label{fig:ann_nw_gr}
\end{center}
\end{figure}

\begin{figure}[htb]
\begin{center}
\includegraphics[scale=0.3,angle=-90]{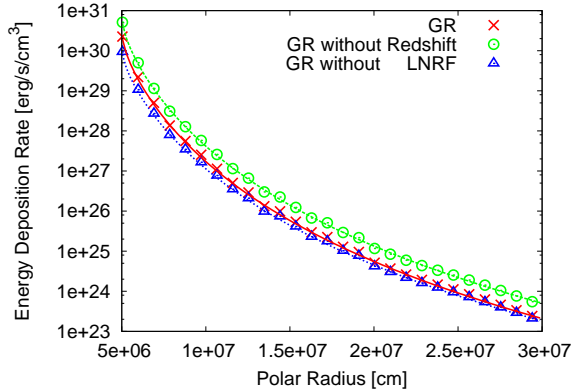}
\caption{Same as Figure \ref{fig:ann_amr} but for 
the numerical solution without the gravitational redshift (circle),
 without the correction due to the tetrad transformation (triangle), and the
 one including both (cross). Agreement with the analytic solution (line) can 
 be seen when the two ingredients are included (See text for detail.)} 
\label{fig:ann_gr_fac}
\end{center}
\end{figure}
 
To find the analytic solution in Equation (\ref{eq:heatdefgr}),
 we first take the most simplest case of $a^* = 0, u^{r} = u^{\theta} = 0$. 
In this case, the momenta $p^{\rm L}_{\alpha}$ in the LNRF can be 
 expressed in the BLF as
\begin{equation}
	p^{\rm L}_{\alpha} = {\bar \omega}^\beta_{~\alpha} p_{\beta }, 
\end{equation}
where
\begin{eqnarray}
{\bar \omega}^\beta_{~\alpha} &\equiv& (\omega^\beta_{~\alpha})^{-1} \nonumber \\
&=& 
\begin{pmatrix}
   \frac{1}{\alpha} & 0 & 0 & 0 \\
   0 & \frac{1}{\sqrt{\gamma_{rr}}} & 0 & 0 \\
   0 & 0 & \frac{1}{\sqrt{\gamma_{\theta \theta}}} & 0\\ 
   -\frac{\beta^{\phi}}{\alpha} & 0 & 0 & \frac{1}{\sqrt{\gamma_{\phi \phi}}} \\
\end{pmatrix}.
\end{eqnarray}
 In this way, the solid angle between the two frames can be readily shown to 
 be the same ($d\Omega^{\rm L} = d\Omega$).
Similarly, the volume element of the phase space and the neutrino energy in 
 the local rest frame can be expressed by the variables in the BLF as follows,
\begin{eqnarray} 
	d^3p^{\rm L} &=& -(p^{\rm L}_0)^2 dp^{\rm L}_0 d\Omega^{\rm L} \nonumber \\
				 &=& -({\bar \omega}^{0}_{0})^3 (p_0)^2  dp_0 d\Omega,
\end{eqnarray}
and
\begin{eqnarray}
	\epsilon^{\rm R}
	&=& -u^\alpha p_{\alpha} \nonumber \\
	&=& -u^0 p_{0},
\end{eqnarray}
where 
\begin{eqnarray}
	u^0 &=& \frac{1}{\sqrt{-g_{\mu \nu} v^\mu v^\nu}} \nonumber \\
	    &=& \frac{1}{\sqrt{-g_{00} - g_{33}(v^3)^2}}, 
\end{eqnarray}
here we define $v^{\mu} = u^{\mu}/u^{0}$. 

Inserting these results to Equation (\ref{eq:heatdefgr}), 
we obtain the following analytic forms of the energy and momentum deposition rate 
respectively as, 
\begin{eqnarray}\label{eq:test_ana}
    Q_{t}^{\rm L} (\mbox{\boldmath$r$}) &=& 
	2 c K G_{\rm F}^2  \xi_\nu^9 (\mbox{\boldmath$r$}) E^{R}_{\nu}(\mbox{\boldmath$r$}) N^{R}_{\nu}(\mbox{\boldmath$r$}) F(\mbox{\boldmath$r$}), \\
	Q_{r}^{\rm L} (\mbox{\boldmath$r$}) &=& 
	2 c K G_{\rm F}^2  \xi_\nu^9 (\mbox{\boldmath$r$}) E^{R}_{\nu}(\mbox{\boldmath$r$}) N^{R}_{\nu}(\mbox{\boldmath$r$}) G(\mbox{\boldmath$r$}).
\end{eqnarray}
 Here $\xi_\nu$ reflects the general relativistic correction to the 
 neutrino energy as 
\begin{eqnarray}
	\xi_{\nu}(\mbox{\boldmath$r$}) &\equiv& 
 \epsilon^{L} / \epsilon^{R}  \nonumber \\ &=& 
	\sqrt{\frac{-g_{00}(R)-g_{33}(v^3)^2}{-g_{00}(r)}} 
	= \sqrt{\frac{1-2M/R-({\hat v}^3)^2}{1-2M/r}},  \label{eq:fac_gr}
\end{eqnarray}
where ${\hat v}^3 = r \sin\theta~v^3$, 
and $R$ is the radius of the neutrino sphere. 
The following two quantities are the energy-weighted integration of the 
 neutrino distribution function on the neutrino sphere 
(namely $f_{\nu}(\mbox{\boldmath$r$}_{\nu}, 
\mbox{\boldmath$p$}_{\nu}^R$)) as,  
\begin{eqnarray}
	E^{R}_{\nu}(\mbox{\boldmath$r$})  &\equiv& \int (\epsilon^{\rm R}_{\nu})^4 f_{\nu}(\mbox{\boldmath$r$}_{\nu}, \mbox{\boldmath$p$}^{\rm R}_{\nu}) d\epsilon^{\rm R}_{\nu} \nonumber \\
	&=& \frac{(kT(\mbox{\boldmath$r$}_\nu))^5}{(hc)^3}{\cal F}_4(0), \\
	N^{R}_{\nu}(\mbox{\boldmath$r$})  &\equiv& \int (\epsilon^{\rm R}_{\nu})^3 f_{\nu}(\mbox{\boldmath$r$}_{\nu}, \mbox{\boldmath$p$}^{R}_{\nu}) d\epsilon^{\rm R}_{\nu} \nonumber \\
	&=& \frac{(kT(\mbox{\boldmath$r$}_\nu))^4}{(hc)^3}{\cal F}_3(0),
\end{eqnarray}
 where $T(\mbox{\boldmath$r$}_\nu)$ is set to be 5 MeV. Finally,
  geometrical factors of $F(r)$ and $G(r)$ are given as,
\begin{eqnarray}
    F(\mbox{\boldmath$r$}) &\equiv& \int \left[ 1-\sin{\theta_{\nu}} \sin{\theta_{\bar{\nu}}}
        \cos{(\varphi_{\nu}-\varphi_{\bar{\nu}})}
        -\cos{\theta_{\nu}} \cos{\theta_{\bar{\nu}}}
    \right]^2 
    d\Omega_{\nu}d\Omega_{\bar{\nu}} \nonumber \\ 
	&=& \frac{2\pi^2}{3} (1-x^4)(5+4x+x^2), \\
    G(\mbox{\boldmath$r$}) &\equiv& \frac{\pi^2}{6} (1-x)^4(1+x)(8+9x+3x^2), \\
	x &\equiv& \sqrt{1-\left(\frac{R}{r}\right)^2 \frac{1-2M/r}{1-2M/R}}. 
\end{eqnarray}

To emphasize the importance of AMR for our ray-tracing calculation (e.g., section 2.2),
 we first show Figure \ref{fig:ann_amr}, in which we compare the energy deposition 
rates calculated with or without AMR treatment. 
 A good agreement with the analytic solution can be obtained by utilizing the AMR 
technique.  We take $n_{\theta} = 32$ with AMR to be the fiducial value 
in the following test calculations.
A visualization of AMR is also given in Figure \ref{fig:amr2d}. 

In Figure \ref{fig:ann_nw_gr}, we compare the analytic solutions (line) 
with the corresponding numerical solutions in the following four cases; 
 the Minkowskian case with or without rotation (circle, cross), 
 the Schwarzschild case with or without rotation (square, triangle).
For models with rotation, we set $1/\sqrt{1- ({\hat v}^3)^{2}} = 2$. 
For models with the Schwarzschild geometry, we put a point mass of $M=3\Ms$ inside 
 the neutrino sphere. It is noted that analytic solutions for each case can be 
readily derived from Equation 
(\ref{eq:test_ana}). In all the cases, the numerical solutions are shown to 
reproduce the corresponding analytic solutions quite well.

Here we present the test calculations to check our implementation of the two GR factors in Equation 
(\ref{eq:fac_gr}), that is the gravitational redshift ($-g_{00}(R)$) and the tetrad transformation ($-g_{00}(r)$). 
On purpose, we neglect each factor one by one, and compare it to the analytical solution. 
Figure \ref{fig:ann_gr_fac} depicts the numerical solutions including 
both (cross) versus without the gravitational redshift (circle) or without 
the tetrad transformation (triangle).
The analytic solutions (lines) are shown to be reproduced only when both of them are appropriately included.



\begin{figure}[tb]
\begin{center}
\includegraphics[scale=0.4]{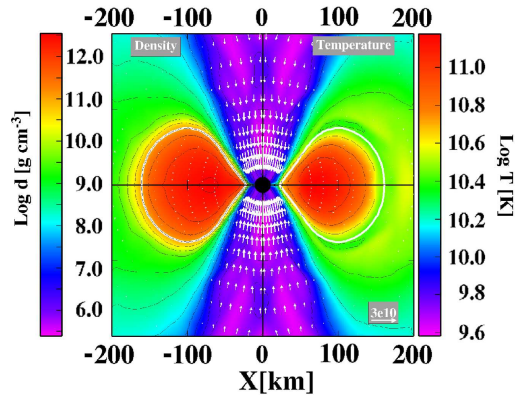}
\caption{Hydrodynamic configuration employed in the ray-tracing calculation.
 This is the snapshot at 9.1 s after the onset of gravitational collapse for 
 model J0.8 when the accretion disk is in a stationary state (see \citet{hari10} 
 for more detail). The logarithmic density (in$\gpcmc$, left-half) and
 temperature (in $K$, right-half) are shown. 
 The white solid line denotes the area where the density is equal to $10^{11} \gpcmc$,
 representing the surface of the accretion disk. 
The central black circle ($\approx 4 M_{\odot}$) represents
the inner boundary of our computations.  
}
\label{fig11a}
\end{center}
\end{figure}

\section{Application to Collapsar Model}
\label{sec:appl}

Having checked the accuracy of our code in previous sections, 
we are now in a position to
 show an application of our code in the collapsar's environment.
As in \citet{hari10}, we estimate the annihilation rates in a post-processing manner 
using the hydrodynamic data obtained in our long-term collapsar simulations.
 By comparing the neutrino-heating timescale to the advection timescale of material 
in the polar funnel regions (see \citet{hari10} for detail),
 we discuss the possibility of generating neutrino-driven outflows there. 
Paying particular attention to the GR effects on the 
 annihilation rates, we discuss their possible impacts on the collapsar dynamics.

As for the hydrodynamic data (such as density, electron fraction, and entropy), 
we take the ones at 9.1 s after the onset of 
gravitational collapse for model J0.8 (Figure \ref{fig11a}), which 
 show a clear accretion-disk and BH system with the polar funnel regions along
 the spin axis of the disk.
 Since this model is calculated by special relativistic 
 hydrodynamics, we project those data into the ones in the LNRF
 for the ray-tracing calculation.
 The position of the inner boundary of 
 the computational domain is set to be $4M_{\odot}$, which mimics 
 the event horizon of the BH. We set the Kerr parameter by 
hand as $a^*=0$ 
for the Schwarzschild geometry and $a^*=0.999$ for the extreme Kerr geometry. 

\clearpage

\begin{figure*}
\begin{center}
\includegraphics[scale=0.4]{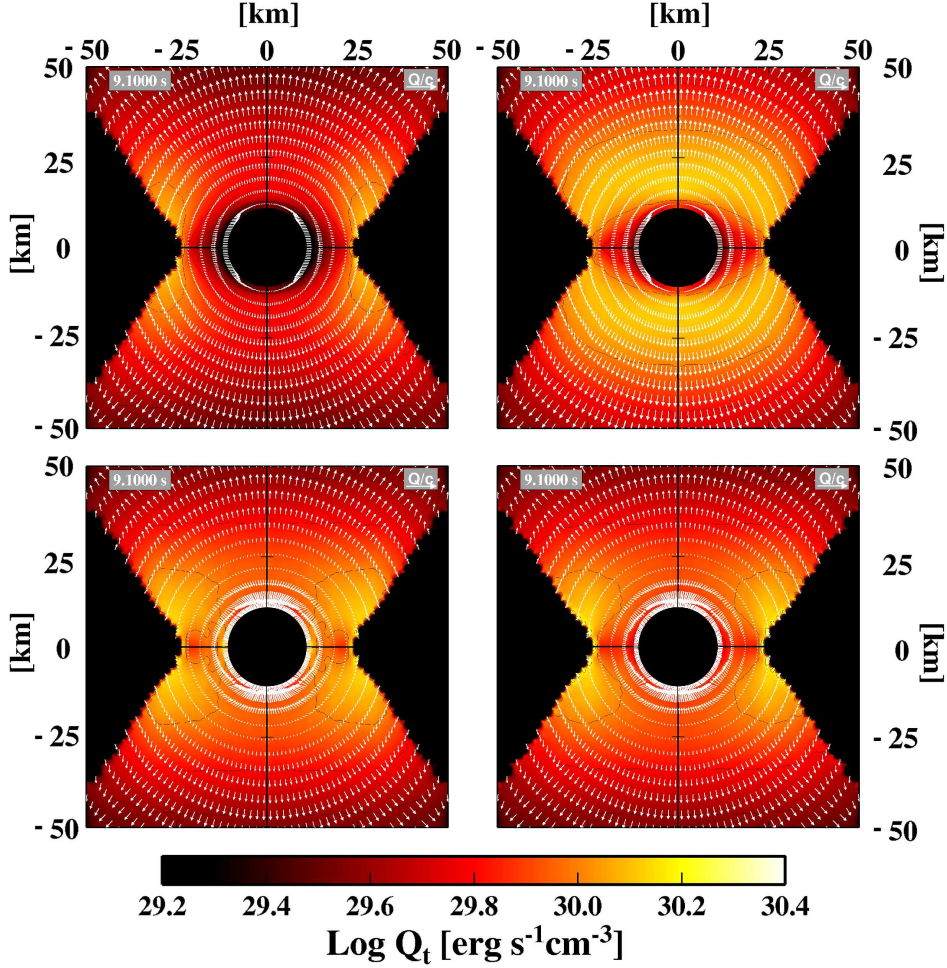}
\caption{Logarithmic contour of energy deposition rates 
$Q_{t}\, [{\rm erg}~{\rm s}^{-1}~{\rm cm}^{-3}]$  and normalized vector 
of momentum deposition rates $Q_{i} / Q_{t}$ calculated for the Minkowskian geometry 
(top left/top right:without/with special relativistic corrections), the 
 Schwarzschild geometry (bottom left), and the extreme Kerr geometry ($a^*=0.999$) 
 (bottom right).  The spatial 
 vector is visualized by showing the spatial velocity vector $v \equiv Q_{i}/Q_{t}$, 
which is normalized by the speed of light $(c=1)$
 being represented by the arrow (top right in each panel).
The central black circle ($\approx 4 M_{\odot}$) 
represents the inner boundary of the computational domain.
Note that the triangular regions colored by
 black closely coincide the surface of the accretion disk (e.g., Figure \ref{fig11a}).}
\label{fig:ann_GR1}
\end{center}
\end{figure*}

\subsection{Effect of General Relativity on Energy and Momentum Deposition}
\label{sec:eff_nu_gr}

\begin{figure}[tbd]
\begin{center}
\includegraphics[scale=0.4]{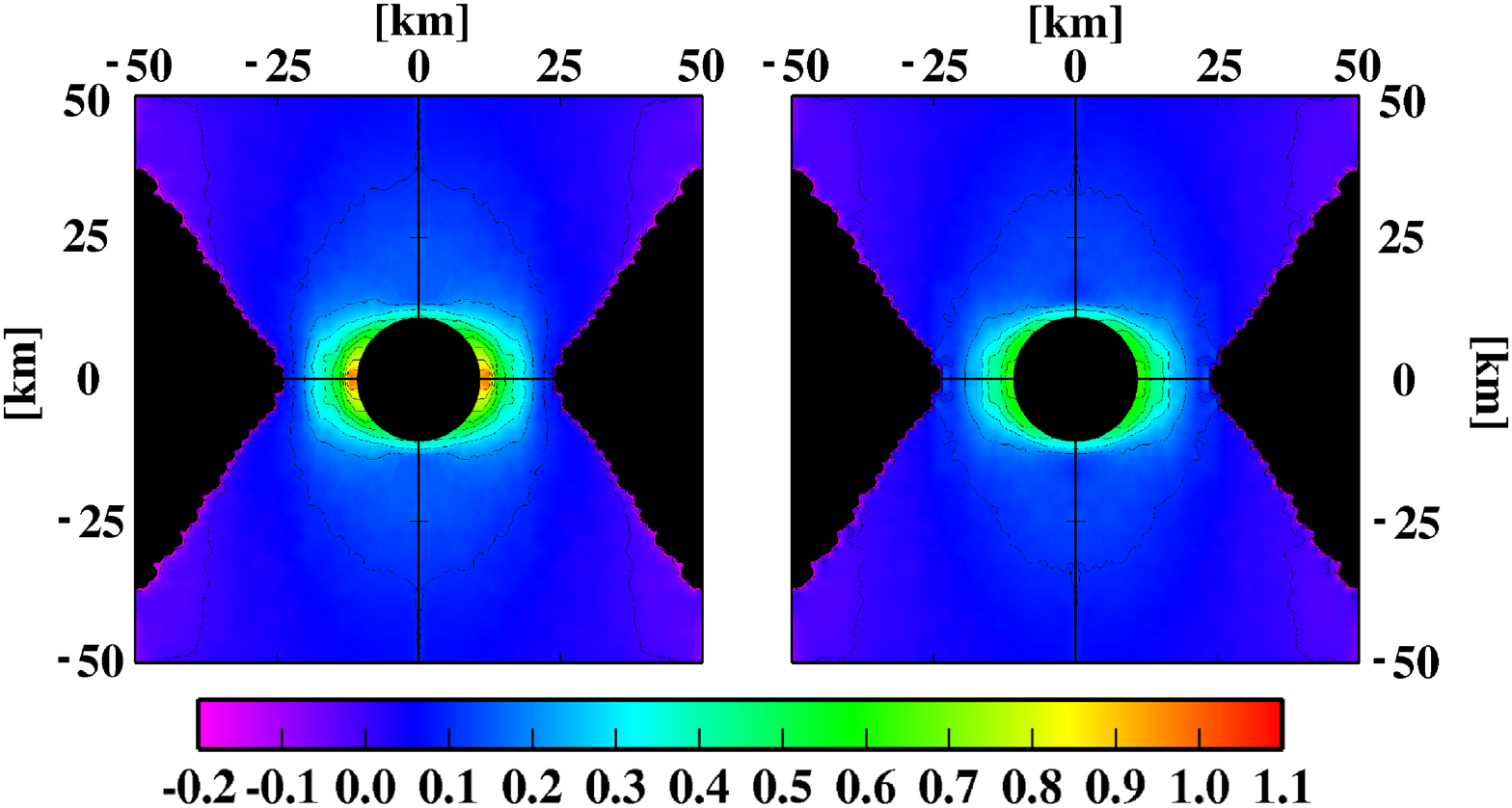}
\caption{Difference plot which shows the 
energy deposition rate in the Schwarzschild (left) and extreme Kerr geometry (right), 
 which are normalized by the one in the Minkowskian geometry 
(see top left in Figure \ref{fig:ann_GR1}).}
\label{fig:ann_GR2}
\end{center}
\end{figure}

To clarify the GR effects, we compare the annihilation rates in 
 the Minkowkian ($M=0$), Schwarzschild, and extreme Kerr geometry ($a^*=0.999$). 
Figure \ref{fig:ann_GR1} shows the energy deposition rate $Q_{t}$ (contour) (Equation \ref{eq:heatdefgr})
and the normalized momentum transfer rate $Q_{i}/Q_{t}$ (vector) 
(e.g., Equation \ref{eq:heatdefgr}) for 
 the Minkowskian geometry (top left/top right: with/without special relativistic 
corrections), 
the Schwarzschild geometry (bottom left), and the extreme Kerr geometry (bottom right), 
respectively. And Figure \ref{fig:ann_GR2} is a difference plot, which shows the 
energy deposition rate normalized by the one in the Minkowskian geometry 
(top left in Figure \ref{fig:ann_GR1}).

From Figure \ref{fig:ann_GR1}, it can be seen in the Minkowskian geometry 
(top two panels) that
 the direction of the momentum transfer is generally radially outward, 
while in the Schwarzschild and Kerr geometry (bottom two panels), 
the direction especially in the vicinity of the BH tends to direct the center 
as a result of the general relativistic bending. 
 This bending effect, acting to suppress the outward momentum transfer,
 should do harm to launch the neutrino-driven outflow. On the other hand,
 it does good to the energy deposition, because it enhances the head-on collision
 especially in the polar funnel regions. 

In fact, it can be seen that the deposition rate 
for the Schwarzschild and Kerr geometry becomes larger than for the Minkowskian 
geometry (Figure \ref{fig:ann_GR2}). In the blueish region that corresponds
 to the polar funnel region, the energy deposition rate for the 
 extreme Kerr geometry is enhanced by factors compared to the Minkowskian geometry. Interestingly it is mentioned that the heating 
 rate is enhanced by about one order-of-magnitude near the equatorial plane in 
the vicinity of the BH. This is because the neutrino rays are concentrated there,
 reflecting the conical shape of the accretion disk (triangular blackish regions 
at the sides). This concentration near the equatorial plane is found to be suppressed 
for the maximally rotating BH mainly due to the frame-dragging effect. 
   

To see the GR effects on the net energy deposition, 
we calculate the total energy deposition rate,
\begin{equation}
Q^{\rm tot}_{\nu {\bar \nu}} = \int \sqrt{-g} Q_{t} dV, 
\label{qtot}
\end{equation}
and the one with the outgoing momentum as \citet{jaros93,birkl}) 
\begin{equation}
Q^{\rm out}_{\nu {\bar \nu}} = \left.\int \sqrt{-g} Q_{t} dV \right|_{Q_{r}> 0},
\label{qtot_out}
\end{equation}
 where the contributions with the outgoing radial component of the momentum vector 
($Q_{r}$) are counted. 
As shown in Table \ref{Table:outflow_gr}, the net deposition rate
and efficiency for the extreme Kerr geometry 
increase up to $16 \%$ ($18 \%$ for 
$Q^{\rm out}_{\nu {\bar \nu}}$) compared to the Minkowskian geometry. 
  Our results support the previous study that GR can enhance the heating rate, and thus 
 good for the formation of the neutrino-driven outflow (e.g., \citet{birkl}).
 From Table 1, the deposition rate and efficiency are barely influenced by the spin of BH. 
 However it should be noted that we have included the spin effects only in the 
radiative transport. As pointed out by \citet{asano1,birkl}, the spin effects, 
such as on the structure of the spacetime (i.e., the inner-most stable circular orbit 
becomes smaller for the rapidly rotating black hole) and also on the accretion disk,
 should be more important to affect the heating rate. 
 To clarify this point, we need a hydrodynamic data based on the general relativistic
 simulations of collapsars, which we are to investigate as a sequel of this study. 

\begin{table}[hbt]
\begin{center}
\begin{tabular}[c]{cccc}
\hline
Geometry		& $Q^{\rm tot}_{\nu {\bar \nu}} [{\rm erg \, \, s^{-1}}]$ & $Q^{\rm out}_{\nu {\bar \nu}}	
[{\rm erg \, \, s^{-1}}]$ & {\rm efficiency} [\%] \\
\hline
Minkowski 		& $6.18\times10^{50}$ & $ 5.71\times10^{50}$ & 0.510 \\
Schwarzschild   & $7.15\times10^{50}$ & $ 6.09\times10^{50}$ & 0.590 \\
Extreme Kerr 	& $7.08\times10^{50}$ & $ 6.13\times10^{50}$ & 0.585 \\
\hline
\end{tabular}
\caption{Comparison of $Q_{\nu {\bar \nu}}^{\rm tot}$, $Q_{\nu {\bar \nu}}^{\rm out}$
 (see Equations (\ref{qtot},\ref{qtot_out})) and efficiency 
for the Minkowskian, Schwarzschild, and extreme 
Kerr ($a^* = 0.999$) geometry, which corresponds to the top left, bottom left and 
right panels in Figure \ref{fig:ann_GR1}, respectively. 
Efficiency is evaluated as $Q^{\rm tot}_{\nu {\bar \nu}}/ L_{\nu}$ where 
$L_{\nu}$ is the total neutrino luminosity. }
\label{Table:outflow_gr}
\end{center}
\end{table}

\clearpage

\begin{figure*}
\begin{center}
\includegraphics[scale=0.5]{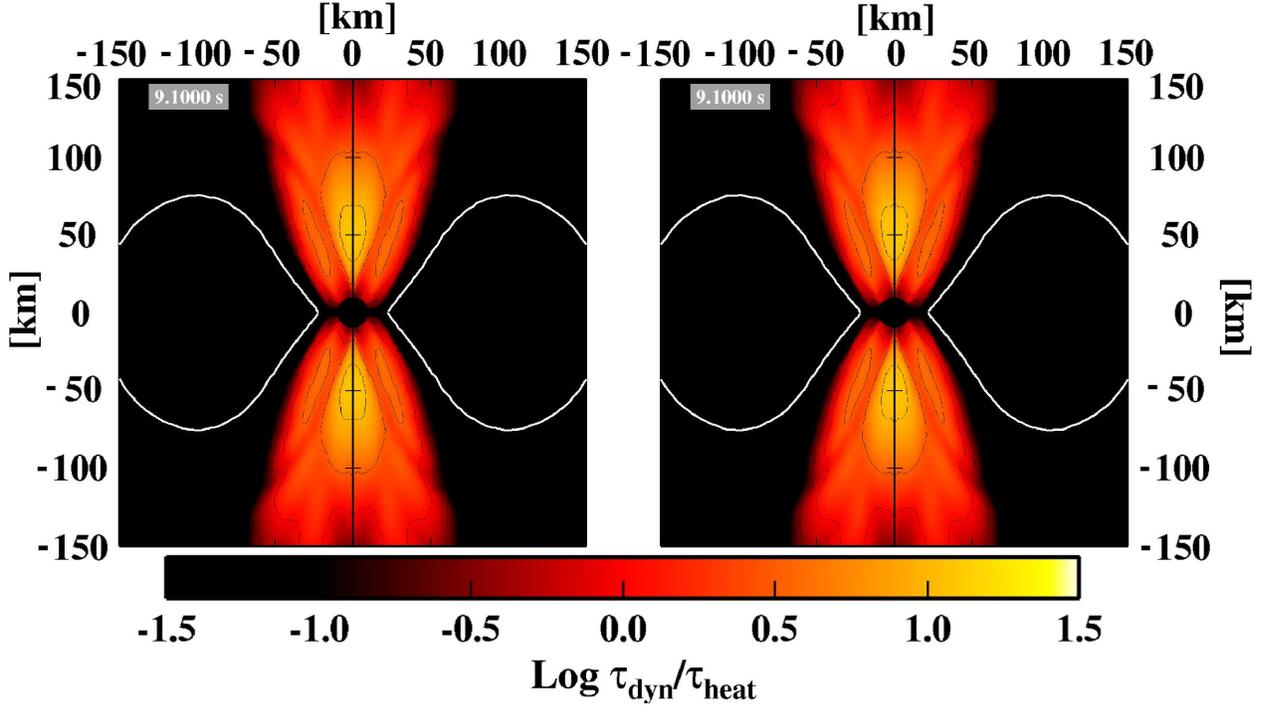}
\caption{Same as Figure \ref{fig:ann_GR1} but for 
$\tau_{\rm dyn} / \tau_{\rm heat}$ (:the dynamical timescale $\tau_{\rm dyn}$ versus the heating timescale 
$\tau_{\rm heat}$)
in the Schwarzschild (left) and extreme Kerr geometry (left), respectively.}
\label{fig:ann_GR3}
\end{center}
\end{figure*}

\subsection{Condition for Outflow Formation}

Based on the annihilation rates in the last section, we compare the two timescales in 
 this section, which are the neutrino-heating timescale and the dynamical timescale. 
Then we anticipate if the neutrino-heating outflows could or could not be produced 
in the polar funnel regions.

To trigger the neutrino-heating explosion, the neutrino-heating timescale 
should be smaller than the advection timescale, which is characterised 
by the free-fall timescale in the polar funnel regions. 
This condition is akin to the condition of the successful neutrino-driven explosion
 in the case of core-collapse supernovae 
(e.g., \citet{bethe} and see collective references in \citet{janka07}).
 The heating timescale is the timescale for a fluid to absorb the energy
 by the neutrino heating, comparable to the gravitational binding energy for making 
 the fluid gravitationally unbound, which may be 
defined as $\tau_{\rm{heat}} \equiv \rho \Phi /Q_{t}$. Here $\Phi$,
 the local gravitational potential, is taken to be the sum of the 
pseudo-Newtonian potential and self-gravity in the flat space-time \citep{hari10} 
and $\rho$ is the local matter density. 
Then the dynamical timescale is defined as 
$\tau_{\rm{dyn}} \equiv \sqrt{3\pi/16G\bar{\rho}}$, 
where $\bar{\rho}$ is the average density at a certain radius and we take 
$\bar{\rho}(r) \equiv 3 M(r) /4 \pi r^3$. 

\begin{figure*}
\begin{center}
\includegraphics[scale=0.6]{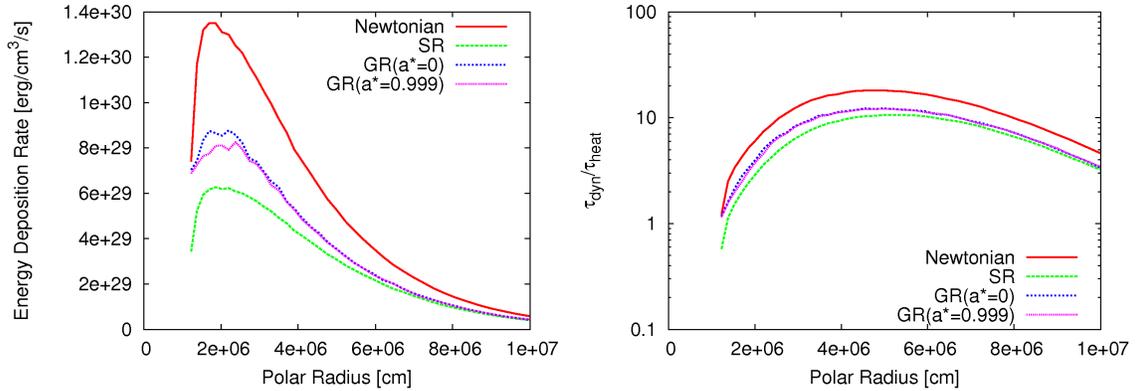}
\caption{Comparison of energy deposition rate (left) and 
$\tau_{\rm dyn} / \tau_{\rm heat}$ along the rotational axis (right)
 between the Minkowskian geometry without or with the special relativistic 
correction (indicated by ``Newtonian'' and ``SR''), and for the Schwarzschild and 
extreme Kerr geometry (indicated by ``GR ($a^*=0$)'' and ``GR ($a^*=0.999$)'').
}
\label{fig:ann_GR4}
\end{center}
\end{figure*}

  Figure \ref{fig:ann_GR3} depicts the ratio of the dynamical $\tau_{\rm{dyn}}$ to 
the heating timescales $\tau_{\rm{heat}}$ for the Schwarzschild (left) 
and extreme Kerr geometry (right), showing in both cases that the ratio becomes 
greater than unity in the polar funnel regions (compare Figure \ref{fig11a}).  
 Figure \ref{fig:ann_GR4} shows the energy deposition rate 
 along the polar axis of Figure \ref{fig:ann_GR1},
for the Minkowskian geometry without or with the special relativistic 
correction (indicated by ``Newtonian'' and ``SR''), and for the Schwarzschild and 
extreme Kerr geometry (indicated by ``GR ($a^*=0$)'' and ``GR ($a^*=0.999$)'').
 It is noted that the energy deposition sharply drops from the 
 Newtonian to the SR case (left panel). This is the outcome of the special 
relativistic beaming effects. Since the rotational velocity of the accretion disk 
is perpendicular to the polar direction, the special relativistic beaming 
effect suppresses the neutrino emission toward the polar region. 
(see \citet{hari10} for more detail). When the 
 general relativistic bending effects are taken into account,
 the deposition rate becomes larger again (see ``GR ($a^*=0$)'' and ``GR ($a^*=0.999$)'').
 Reflecting this situation, $\tau_{\rm dyn} / \tau_{\rm heat}$ becomes 
 smallest for the case with SR and largest for the Newtonian case (right panel 
 of Figure \ref{fig:ann_GR4}). It is important that the ratio in 
 the case of the Schwarzschild and extreme Kerr geometry, which do reflect nature
 in the collapsar's environment, 
becomes larger than unity inside 100 km in the vicinity of the rotational axis
 (Figure \ref{fig:ann_GR4}, right).
This indicates the possible formation of the
 neutrino-driven outflows there, if coupled to the collapsar's hydrodynamics.

\section{Summary and Discussion}
\label{sec:discussion}

In the light of collapsar models of gamma-ray bursts (GRBs), we developed 
a numerical scheme and code for estimating the deposition of energy and 
momentum due to the neutrino pair annihilation 
($\nu + {\bar \nu} \rightarrow e^{-} + e^{+}$) in the vicinity of accretion tori
 around a Kerr black hole.
We designed our code to calculate the general relativistic neutrino transfer 
 by a ray-tracing method.  To solve the collisional Boltzmann 
equation in the Kerr geometry, we numerically 
integrated the so-called rendering equation along the null geodesics. 
For the neutrino opacity, the charged-current $\beta$-processes are taken into 
account, which are dominant in the vicinity of the accretion tori.
  We employed the Fehlberg(4,5) adaptive integrator in the Runge-Kutta method
 in order to perform the numerical integration accurately.
 We checked the numerical accuracy of the developed code by several
 tests, in which we showed comparisons with the corresponding analytical solutions. 
 In order to solve the energy dependent ray-tracing transport, 
we proposed that an adaptive-mesh-refinement approach, which we took
for the two radiation angles $(\theta,\phi)$ and the neutrino energy, is efficient to 
 reduce the computational cost.  Based on the hydrodynamical data in our 
collapsar simulation, we estimated the annihilation rates 
in a post-processing manner.
 It is found that 
the general relativistic effect can increase the 
local energy deposition rate by about one order of magnitude, and 
the net energy deposition rate by several tens of percents.
After the accretion disk settles into a stationary state (typically later than 
 $\sim 9$ s from the onset of gravitational collapse),
 we pointed out that the neutrino-heating timescale 
can be smaller than the dynamical timescale inside 100 km in the vicinity of the 
rotational axis. Our results 
suggest that the neutrino-driven outflows  can possibly be launched there.

For further investigation, we need to include several important ingredients 
 ignored in this study. We plan to develop a GRMHD code for collapsars, 
which is indispensable to see the outcome of this paper.
 By changing the precollapse magnetic fields and rotation systematically,
 we hope to clearly understand how the outflow formation in collapsars could 
change from the neutrino-driven mechanism to the MHD-driven one. The neutrino 
 oscillation by the Mikheyev-Smirnov-Wolfenstein (MSW) effect 
 (see collective references in \citet{kotake_rev,kawa09}) could be important, 
albeit in much later phase than we considered in this paper.
 When the density in the polar funnel regions drops as low as 
$\rho \lesssim 10^{3} \gpcmc$ later, the neutrino oscillation could operate for 
neutrinos traveling from the accretion disk to the polar funnel. If this is the 
case, the incoming neutrino spectra to the polar funnel regions and the 
pair annihilation rates there could be affected significantly. 
 It is also noted that the effects of neutrino self-interaction are 
 remained to be studied, which has been attracting great attention 
 in the theory of core-collapse supernovae (e.g., \citet{duan06}). 
As in the case of core-collapse supernovae (e.g., \citet{kotake_apjl,ott_rev}), 
studies of gravitational-wave emissions from 
 collapsars might provide us a new window to probe into the central engine 
(e.g., \citet{hira_05,suwa_09}).
As a sequel of this work, we are planning to implement the ray-tracing calculation 
 to the GRMHD simulation and clarify these issues one by one. 
 We hope that this study takes a very first step towards 
 the meeting of GR with neutrino transport, which should be 
indispensable for understanding the collapsar engines.

\acknowledgements{
S.H. is grateful to T. Kajino for helpful exchanges.
T.T. and K.K. express thanks to K. Sato, S. Yamada, and S. Nagataki
 for continuing encouragements.
Numerical computations were in part carried on XT4 and 
general common use computer system at the center for Computational Astrophysics, CfCA, the National Astronomical Observatory of Japan.  This
study was supported in part by the Grants-in-Aid for the Scientific Research 
from the Ministry of Education, Science and Culture of Japan
(Nos. S19104006, 19540309 and 20740150).

\bibliographystyle{apj}
\bibliography{ms}

\end{document}